# Chiral Dynamics and Single-Spin Asymmetries

Dennis Sivers

Portland Physics Inst.
4730 SW Macadam, #101
Portland, OR 97239

Spin Physics Center
University of Michigan
Ann Arbor, MI 48109

ABSTRACT

Parity-conserving single-spin asymmetries provide a specific measure of coherent spin-orbit dynamics in quantum chromodynamics. The origin of these effects can be traced to the interplay of chiral dynamics and confinement in the theory. The most elegant display of the relevant mechanisms occurs in the Collins functions and in the polarizing fragmentation functions and fracture functions for particles with spin. In the nucleon, these same dynamical mechanisms generate virtual quantum structures leading to the Boer-Mulders functions and orbital distributions. Two complementary formalisms for these distribution functions appear. The familiar gauge-link formalism incorporates all nonperturbative dynamics into nonlocal correlators. The <u>constructive</u> formalism introduced by the author describes distributions normalized to an intrinsic property of the nucleon, namely, the currents specified in the Bakker-Leader-Trueman sum rule. The connection between these two approaches can be explored in the process dependence of single-spin asymmetries in various hard-scattering processes. The study of the SU(2) Weyl-Dirac equation in spherical coordinates allows typical Wilson operators that determine this process dependence to be evaluated in the coordinate gauge.

This conference has already heard two excellent theoretical talks on transverse single-spin asymmetries by Professors Efremov [1] and Teryaev [2]. A specific goal of this presentation is to acquaint you with a different, and complementary, set of theoretical tools that emphasizes the dynamical origin of these observables. A convenient way of introducing this alternative approach focuses on the concept of spin-directed momentum. The observation that single-spin measurements (either analyzing powers or polarizations) necessarily define a spin-directed momentum can be easily confirmed. The requirement can be illustrated by the sketch shown in Fig. 1. For a parity-conserving asymmetry, the form of this spin-directed momentum is highly constrained by rotational invariance and finite symmetries. The required expression is

$$k_{TN} = \vec{k}_T \cdot (\hat{\sigma} \times \hat{P}) \tag{1.1}$$

in which $\vec{P}$ is the 3-momentum of a hadron, $\vec{k}$ the 3-momentum of a constituent and $\vec{\sigma}$ and axial vector denoting a spin direction. This expression can easily seen to be invariant under C (charge conjugation), P (parity) and T (time reflection). It is, however, odd under a symmetry designated $A_\tau$ [3,4]. The observation that all single-spin observables are odd under the combination $O = PA_\tau$ requires that all such observables fall into one of two distinct categories:
   1. P-odd and $A_\tau$-even,
   2. P-even and $A_\tau$-odd.

In the light-quark sector of the standard model, $A_\tau$-odd observables can be shown to be uniquely associated with coherent spin-orbit dynamics. Mulders and Tangerman [5] have classified four distinct leading-twist functions characterizing $A_\tau$-odd quantum structures. This collection consists of two types of fragmentation function: the Collins functions [6] and the polarizing fragmentation functions, and two types of distribution function; the Boer-Mulders functions [7] and the orbital distribution functions. These functions have distinct characteristics but they share a common origin in the combination of confinement and chiral dynamics that generate the non-perturbative spin-orbit correlations.

The existence of a probabilistic description of the $A_\tau$-odd dynamics in fragmentation functions is guaranteed by the existence of a projection operator [3,4]

$$P_A^- = \frac{1 - A_\tau}{2} \tag{1.2}$$

that isolates spin-orbit effects. The work of Artru, Czyzewski, and Yabuki [8] displays these dynamical elements very elegantly. This model actually has all the ingredients of the full nonperturbative calculation of the pion Collins function. The ingredients include mixing between gluonic degrees of freedom and a $0^{++}$ $^3P_0$ quark-antiquark pair that generates internal orbital angular momentum. As the pair rotates, configuration mixing alters the local SU(3) color geometry, enhancing the probability of flux-tube breaking. Chiral dynamics enter the picture by giving an energy advantage for the antiquark in the

$^3P_0$ pair to form a light-mass pion involving the leading quark. This pion then inherits the spin-directed momentum of the antiquark. The phenomenological estimates for pion Collins functions presented in Prof. Efremov's talk [1] provide strong quantitative support for this picture. The reader should consult his summary for the original references.

For particles with spin, a density matrix formulation of these basic dynamical mechanisms with a quantization axis specified by the orientation of $\vec{L}$ produces a tightly-constrained formulation of both the Collins function and the polarizing fragmentation function. For baryons, it is convenient to consider, in addition to usual polarizing fragmentation functions describing the fragmentation of a quark jet, the polarizing fracture function describing the fragmentation of the "diquark-jet" created by stripping away a quark from a nucleon target to create an SU(3) color ion. This extension of the fracture-function formalism established by Trendadue and Veneziano [9] to the sector of $A_\tau$-odd dynamics seems also to be efficiently characterized by mechanisms similar to those of the Collins functions.

Since the nucleon is a stable particle, the orbital angular momentum that appears explicitly in the final state for fragmentation and fracture functions appears as virtual quantum structures leading to the nucleon's $A_\tau$-odd distribution functions. The field-theoretical descriptions of the virtual processes allow for two distinct formalisms to characterize these distributions. The distinction between the two approaches reflects the alternate descriptions for single-spin observables illustrated in Fig. 1. The, now conventional, gauge-link formalism [10,11] presents these functions in terms of nonlocal correlators that lead to the expectation value for $k_{TN}$. This formalism engages the full power of gauge theory and makes a direct connection to the operator product expansion. [12]. The predictive power of the gauge-link approach is demonstrated by the Collins conjugation relation, [10] that relates the orbital distribution measured in SIDIS with that measured in the DY process. The other, more modest, formalism, developed by the author in [3,4], is based on local, gauge-invariant, number densities that describe properties intrinsic to the proton and are not based on any specific process. The orbital distributions and Boer-Mulders functions in this formalism are constructed such that they are normalized to the expectation values of spin-orbit effects,

$$\int dx d^2 k_T \Delta^N G_{q/p\uparrow}^{front}(x, k_{TN}(x); \mu^2) = \frac{1}{2} \left\langle \vec{L}_q \cdot \hat{\sigma}_p(\mu^2) \right\rangle$$

$$\int dx d^2 k_T \Delta^N G_{g/p\uparrow}^{front}(x, k_{TN}(x); \mu^2) = \frac{1}{2} \left\langle \vec{J}_g \cdot \hat{\sigma}_p(\mu^2) \right\rangle \quad (1.3)$$

$$\int dx d^2 k_T \Delta^N G_{q\uparrow/p}^{front}(x, k_{TN}(x); \mu^2) = \frac{1}{2} \left\langle \vec{L}_q \cdot \hat{\sigma}_q(\mu^2) \right\rangle$$

To avoid confusion, the symbols for the distributions in these constructions are purposely chosen to be different from the Mulders-Tangerman symbols traditionally used in the gauge-link formalism. This, constructive, formalism describes characteristics of the spin-orbit dynamics of the proton so that, for example, the Bakker, Leader, Trueman sum rule [13] can be written,

$$J_y = \frac{1}{2} = \frac{1}{2}\sum_{q_i}\delta^T q_i(\mu^2) + 2\sum_{q_i,g=c}\int dx d^2k_T \Delta^N G^{front}_{c/p\uparrow}(x, k_{TN}(x); \mu^2) \tag{1.4}$$

where $\delta^T q_i(\mu^2)$ is the moment of the quark transversity distribution. The constructive formalism takes advantage of the fact that all $A_\tau$-odd dynamics can be factorized into an effective distribution to give a recipe for the initial-state and final-state interactions that contribute to a given single-spin asymmetry. The specific construction of these functions described in refs. [3,4] clarifies the distinction between constructive distributions and the conventional functions specified by nonlocal correlators.

As discussed in the presentation by Teryaev, [2] at this conference and emphasized, for example by Brodsky [14] the orbital distributions and Boer-Mulders functions defined in the gauge-link formalism are, in fact, effective distributions. These functions are connected to the imaginary parts generated in the helicity-amplitude basis by soft initial-state and/or final-state interactions. In the gauge-link formalism it thus makes perfect sense to say that the distributions are <u>created</u> by the initial-state or final-state interactions involved. Whereas, in the constructive approach of the author it is more correct to say that the underlying $A_\tau$-odd number densities are <u>revealed</u> by the soft initial-state or final-state interactions. Brodsky and his collaborators have demonstrated in numerous calculations [14] the connection between in soft initial-state and final-state interactions involved in single-spin effects with those that appear in other phenomenological contexts.

The constructive approach allows a more direct connection of the virtual corrections leading to quantum structures in the nucleon with the explicit spin-orbit dynamics as displayed by the Collins functions. For example, the calculations presented in Ref. 4 provide the normalization for the quark, antiquark and gluon orbital distributions $\Delta^N G^{front}_{c/p\uparrow}$ and for the quark Boer-Mulders functions $\Delta^N G^{front}_{q\uparrow/p}$ in terms of the expectation values for orbital angular momentum found in the Georgi-Manohar [15] chiral quark model. This venerable model is defined in terms of transitions from constituent quarks, (U,D) to partonic quarks (u,d,s) as exemplified by

$$U\uparrow \to [1-\eta_B - \alpha_c(1+\varepsilon_s+\varepsilon_o)]u\uparrow ...(L=0)$$
$$+[\eta_B u\downarrow +\alpha_c d\downarrow(\bar{d}u)+\varepsilon_s\alpha_c s\downarrow(\bar{s}u)+\varepsilon_o\alpha_c u\downarrow(\bar{u}u)]...(L=+1) \tag{1.5}$$

The transitions for $U\downarrow, D\uparrow$ and $D\downarrow$ can be obtained from (1.5) using isospin and rotational invariance. The fixing of the parameters $\eta_B, \alpha_c, \varepsilon_s, \varepsilon_o$ in the model is an interesting exploration of angular momentum sum rules. The observation that the transitions (1.5) are precisely the Collins functions $q_i\uparrow \to q_j\downarrow \pi_{\bar{j}i}$ emphasizes the underlying connections.

The familiar gauge-link formalism of the effective approach and the transversity-amplitude based calculations of the constructive approach are therefore seen to be very complementary. Each constrains the other in many important ways. Comparing the two leads to a challenging program of study of the process-dependence in single-spin asymmetries. The process dependence for the gauge-link formalism can be described recursively using spectator models and the twist expansion. A convenient tool that goes beyond perturbation theory for beginning to understand the same process dependence in the constructive approach uses explicit solutions of the Weyl-Dirac equation in color SU(2) and spherical coordinates to calculate the Wilson operators that appear for these observables in the coordinate gauge. The techniques for doing these calculations exploit specific operators discussed in Refs.[16,17]. The Weyl-Dirac equation allows the separation of degrees of freedom for the energy and 3-momentum and the coordinate gauge in spherical coordinates simplifies the calculation of the spatial component of Wilson operators as indicated in Fig. 2.

Significant progress is being made in the understanding of the factorization properties [18] of the gauge-link formalism, as well as the connection to the twist expansion [19] and in the relationships with generalized parton distributions [20] found in this approach. The application of the $A_\tau$ symmetry and the formulation of single- spin observables in terms of transversity amplitudes provides important constraints in all these endeavors. The ability to formulate calculations in complementary formalisms has already proven to be a real benefit but the important information contained in the fragmentation and fracture functions has not yet been fully exploited.

**REFERENCES**

1. A.V. Efremov, Invited talk, this conference
2. O.V. Teryaev, Invited talk, this conference
3. D. Sivers, "Single-Spin Observables and Orbital Structures in Hadronic Distributions", ArXiv hep-ph/060908, Phys. Rev.**D74**, 094008(2006)erratum **D75** e039901
4. D. Sivers, "Chiral Mechanisms Leading to Orbital Quantum Structures in the Nucleon", ArXiv 074.1791 (2007) Submitted to Phys Rev D but rejected
5. P.J. Mulders and R.D. Tangerman, Nucl. Phys. **B461**, 197 (1996)
6. J.C. Collins, Nucl. Phys. **B396**, 161 (1993)
7. D. Boer and P.J. Mulders, Phys. Rev. **D57**, 5780 (1994)
8. A. Artru, J. Czyzewski and H. Yabuki, Z. Phys **C73**, 527 (1997)
9. L. Trentadue and G. Veneziano, Phys. Lett. **B323**, 201 (1994)
10. J.C. Collins, Phys. Lett. **B536**, 43 (2002)
11. A.V. Belitsky, X. Ji, and F. Yuan, Nucl. Phys. **B636**, 165 (2003); D. Boer, P.J. Mulders and F. Pijlman, Nucl. Phys. **B667**, 201 (2003)
12. Two excellent comprehensive summaries of the gauge-link formalism can be found in F. Pijlman, Free University of Amsterdam thesis (2006) and C.J. Bonhoff, Free University of Amsterdam thesis (2007
13. B.L. Bakker, E. Leader and T.L. Trueman, Phys. Rev. **D70**, 114001 (2004)

14. See, for example, S.J. Brodsky, ArXiv 0709.2229 (2007) and references contained therein.
15. A. Manohar and H. Georgi, Nucl Phys. **B234**, 189 (1984)
16. J. Ralston and D. Sivers, Phys. Rev. **D28**, 953 (1983); ibid. **D30**, 472 (1984)
17. D. Sivers, Phys. Rev. **D35**, 707 (1987); ibid. **D35**, 3231 (1987)
18. A. Bacchetta, C. Bomhof, U. D'Alesio, P. Mulders and F. Murgia, ArXiv 0703.153
19. P. Ratcliffe and O. Teryaev, arXiv:hep-ph/0703293 (2007)
20. S. Meissner, A. Metz and K. Goeke, arXiv:0706.1193

**Figure Captions**

Fig. 1     Graphical illustration of how a measurement of $A_N$ for a parity-conserving single-spin asymmetry can also be used to define the underlying spin-directed momentum transfer in the process.

Fig. 2     In spherical coordinates, the coordinate gauge allows for the calculation Wilson operators consisting of triangles with two radially directed lines based on the operator techniques discussed in references 16 and 17.

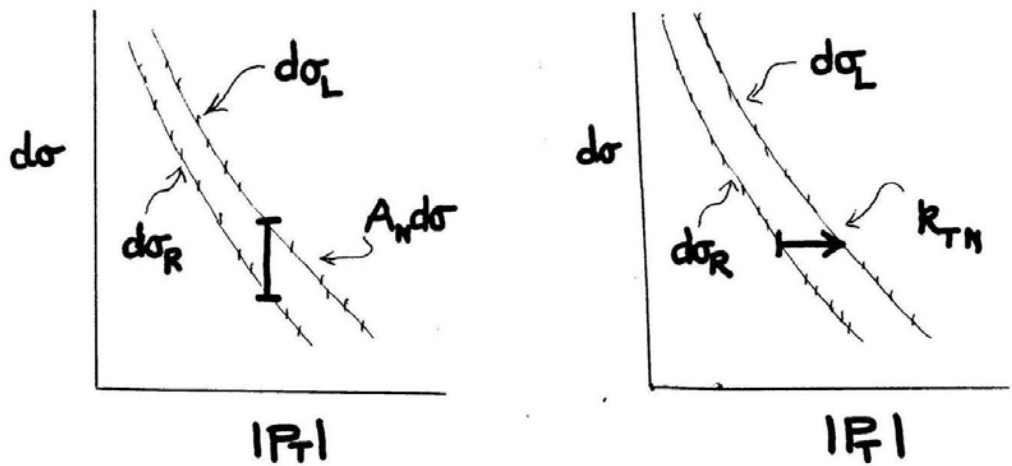

Fig. 1

Spin-Directed Momentum

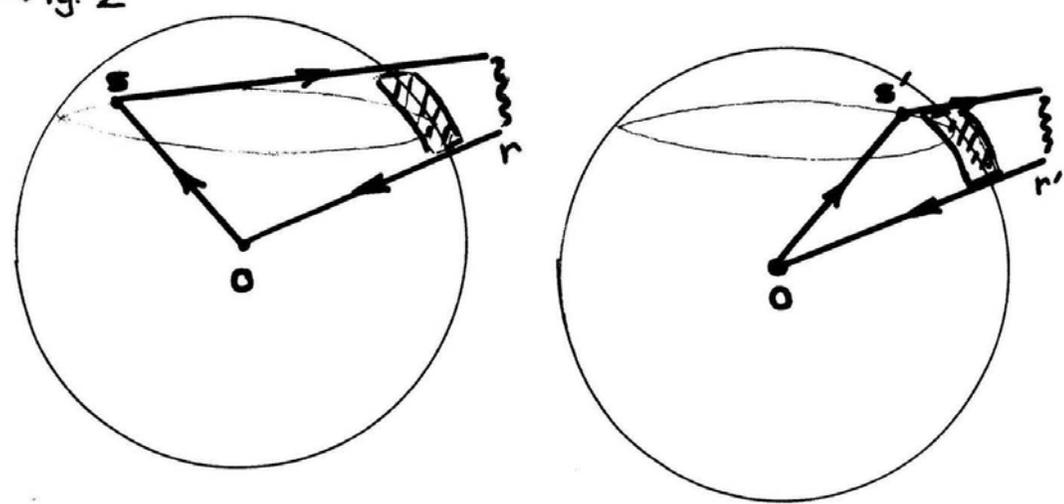

Fig. 2

Wilson Operators